\documentclass[%
 reprint,
 superscriptaddress,
 amsmath,amssymb,
 aps,
 prc,
nolongbibliography,
floatfix,
]{revtex4-1}
\usepackage[utf8]{inputenc}
\usepackage{graphicx}
\usepackage{siunitx}
\usepackage{tabularx}
\DeclareSIUnit\clight{\text{\ensuremath{c}}} 
\usepackage{feynmf}

\begin{document}


\title{Low-$Q^2$ elastic electron-proton scattering using a gas jet target}

\author{Y.~Wang}
\affiliation{Laboratory for Nuclear Science, Massachusetts Institute of Technology, Cambridge, Massachusetts 02139, USA}

\author{J.C.~Bernauer}%
 \email{jan.bernauer@stonybrook.edu}
\affiliation{Center for Frontiers in Nuclear Science, Department of Physics and Astronomy, Stony Brook University, New York 11794, USA}
\affiliation{RIKEN BNL Research Center,  Brookhaven National Laboratory, Upton, NY 11973, USA}

\author{B.S.~Schlimme}
\affiliation{Institut f\"ur Kernphysik, Johannes Gutenberg-Universit\"at, D-55099 Mainz, Germany}

\author{P.~Achenbach}
\author{S.~Aulenbacher}
\affiliation{Institut f\"ur Kernphysik, Johannes Gutenberg-Universit\"at, D-55099 Mainz, Germany}

\author{M.~Ball}
\affiliation{Helmholtz-Institut f\"ur Strahlen- und Kernphysik, Rheinische Friedrich-Wilhelms-Universit\"at, D-53115 Bonn, Germany}  

\author{M.~Biroth}
\affiliation{Institut f\"ur Kernphysik, Johannes Gutenberg-Universit\"at, D-55099 Mainz, Germany}

\author{D.~Bonaventura}
\affiliation{Institut f\"ur Kernphysik, Westf\"alische Wilhelms-Universit\"at, D-48149 M\"unster, Germany}

\author{D.~Bosnar}
\affiliation{Department of Physics, Faculty of Science, University of Zagreb, 10000 Zagreb, Croatia}

\author{P.~Brand}
\affiliation{Institut f\"ur Kernphysik, Westf\"alische Wilhelms-Universit\"at, D-48149 M\"unster, Germany}

\author{S.~Caiazza} 
\author{M.~Christmann} 
\affiliation{Institut f\"ur Kernphysik, Johannes Gutenberg-Universit\"at, D-55099 Mainz, Germany}

\author{E.~Cline} 
\affiliation{Center for Frontiers in Nuclear Science, Department of Physics and Astronomy, Stony Brook University, New York 11794, USA}

\author{A.~Denig} 
\affiliation{Institut f\"ur Kernphysik, Johannes Gutenberg-Universit\"at, D-55099 Mainz, Germany}
\affiliation{PRISMA$^+$ Cluster of Excellence, Johannes Gutenberg-Universit\"at, D-55099 Mainz, Germany}
\affiliation{Helmholtz Institute Mainz, GSI Helmholtzzentrum für Schwerionenforschung, Darmstadt, Johannes Gutenberg-Universit\"at, D-55099 Mainz, Germany}

\author{M.O.~Distler}
\affiliation{Institut f\"ur Kernphysik, Johannes Gutenberg-Universit\"at, D-55099 Mainz, Germany}

\author{L.~Doria}
\affiliation{Institut f\"ur Kernphysik, Johannes Gutenberg-Universit\"at, D-55099 Mainz, Germany}
\affiliation{PRISMA$^+$ Cluster of Excellence, Johannes Gutenberg-Universit\"at, D-55099 Mainz, Germany}

\author{P.~Eckert}
\author{A.~Esser}
\affiliation{Institut f\"ur Kernphysik, Johannes Gutenberg-Universit\"at, D-55099 Mainz, Germany}

\author{I.~Fri\v{s}\v{c}i\'c}
\affiliation{Laboratory for Nuclear Science, Massachusetts Institute of Technology, Cambridge, Massachusetts 02139, USA}

\author{S.~Gagneur}
\author{J.~Geimer} 
\affiliation{Institut f\"ur Kernphysik, Johannes Gutenberg-Universit\"at, D-55099 Mainz, Germany}

\author{S.~Grieser} 
\affiliation{Institut f\"ur Kernphysik, Westf\"alische Wilhelms-Universit\"at, D-48149 M\"unster, Germany}

\author{P.~G\"ulker}
\author{P.~Herrmann}
\author{M.~Hoek}
\author{S.~Kegel}
\affiliation{Institut f\"ur Kernphysik, Johannes Gutenberg-Universit\"at, D-55099 Mainz, Germany}

\author{J.~Kelsey}
\affiliation{MIT Bates Research and Engineering Center, Middleton, Massachusetts, 01949, USA}

\author{P.~Klag}
\affiliation{Institut f\"ur Kernphysik, Johannes Gutenberg-Universit\"at, D-55099 Mainz, Germany}

\author{A.~Khoukaz} 
\affiliation{Institut f\"ur Kernphysik, Westf\"alische Wilhelms-Universit\"at, D-48149 M\"unster, Germany}

\author{M.~Kohl} 
\affiliation{Department of Physics, Hampton University, Hampton, Virginia 23668, USA}

\author{T.~Kolar}
\affiliation{Jo\v{z}ef Stefan Institute, SI-1000 Ljubljana, Slovenia}

\author{M.~Lau{\ss}}
\affiliation{Institut f\"ur Kernphysik, Johannes Gutenberg-Universit\"at, D-55099 Mainz, Germany}

\author{L.~Le{\ss}mann}
\affiliation{Institut f\"ur Kernphysik, Westf\"alische Wilhelms-Universit\"at, D-48149 M\"unster, Germany}

\author{M.~Littich}
\author{S.~Lunkenheimer}
\affiliation{Institut f\"ur Kernphysik, Johannes Gutenberg-Universit\"at, D-55099 Mainz, Germany}

\author{J.~Marekovi\v{c}}
\affiliation{Jo\v{z}ef Stefan Institute, SI-1000 Ljubljana, Slovenia}

\author{D.~Markus} 
\affiliation{Institut f\"ur Kernphysik, Johannes Gutenberg-Universit\"at, D-55099 Mainz, Germany}

\author{M.~Mauch} 
\affiliation{Helmholtz Institute Mainz, GSI Helmholtzzentrum für Schwerionenforschung, Darmstadt, Johannes Gutenberg-Universit\"at, D-55099 Mainz, Germany}

\author{H.~Merkel} 
\affiliation{Institut f\"ur Kernphysik, Johannes Gutenberg-Universit\"at, D-55099 Mainz, Germany}
\affiliation{PRISMA$^+$ Cluster of Excellence, Johannes Gutenberg-Universit\"at, D-55099 Mainz, Germany}

\author{M.~Mihovilovi\v{c}} 
\affiliation{Institut f\"ur Kernphysik, Johannes Gutenberg-Universit\"at, D-55099 Mainz, Germany}
\affiliation{Faculty of Mathematics and Physics, University of Ljubljana, SI-1000 Ljubljana, Slovenia}
\affiliation{Jo\v{z}ef Stefan Institute, SI-1000 Ljubljana, Slovenia}

\author{R.G.~Milner}
\affiliation{Laboratory for Nuclear Science, Massachusetts Institute of Technology, Cambridge, Massachusetts 02139, USA}

\author{J.~M\"uller}
\author{U.~M\"uller}
\affiliation{Institut f\"ur Kernphysik, Johannes Gutenberg-Universit\"at, D-55099 Mainz, Germany}

\author{T.~Petrovi\v{c}}
\affiliation{Jo\v{z}ef Stefan Institute, SI-1000 Ljubljana, Slovenia}

\author{J.~Pochodzalla}
\author{J.~Rausch}
\author{J.~Schlaadt}
\author{H.~Sch\"urg}
\affiliation{Institut f\"ur Kernphysik, Johannes Gutenberg-Universit\"at, D-55099 Mainz, Germany}

\author{C.~Sfienti}
\affiliation{Institut f\"ur Kernphysik, Johannes Gutenberg-Universit\"at, D-55099 Mainz, Germany}
\affiliation{PRISMA$^+$ Cluster of Excellence, Johannes Gutenberg-Universit\"at, D-55099 Mainz, Germany}

\author{S.~\v{S}irca}
\affiliation{Faculty of Mathematics and Physics, University of Ljubljana, SI-1000 Ljubljana, Slovenia}
\affiliation{Jo\v{z}ef Stefan Institute, SI-1000 Ljubljana, Slovenia}

\author{R.~Spreckels} 
\author{S.~Stengel}
\author{Y.~St\"ottinger} 
\author{C.~Szyszka}
\author{M.~Thiel}
\affiliation{Institut f\"ur Kernphysik, Johannes Gutenberg-Universit\"at, D-55099 Mainz, Germany}

\author{S.~Vestrick}
\affiliation{Institut f\"ur Kernphysik, Westf\"alische Wilhelms-Universit\"at, D-48149 M\"unster, Germany}

\author{C.~Vidal}
\affiliation{MIT Bates Research and Engineering Center, Middleton, Massachusetts, 01949, USA}

\collaboration{for the A1 and MAGIX Collaborations}

\date{\today}

\begin{abstract}
In this paper, we describe an experiment measuring low-$Q^2$ elastic electron-proton scattering using a newly developed cryogenic supersonic gas jet target in the A1 three-spectrometer facility at the Mainz Microtron. We measured the proton electric form factor within the four-momentum transfer range of $0.01\le Q^2 \le $ \SI{0.045}{(GeV/\clight)^2}. The experiment showed consistent results with the existing measurements. The data we collected demonstrated the feasibility of the gas jet target and the potential of future scattering experiments using high-resolution spectrometers with this gas jet target.

\end{abstract}
\maketitle
\section{Introduction}

The proton's electric form factor is one of its fundamental properties, reflecting the distribution of charge inside the proton. The proton's size, expressed by the charge radius, is directly related to the form factor slope at zero four-momentum transfer. 

The last decade saw the rise of the proton radius puzzle. Besides the determination via the form factor slope, the radius can be determined also via the hydrogen fine structure energy level splittings from electronic hydrogen atom spectroscopy. The results from these two methods agreed with each other consistently over the years on the value of \SI{0.8775\pm0.0051}{fm} recommended by CODATA2010~\cite{CODATA2010}. However, in 2010, Pohl \emph{et al.} from the CREMA collaboration measured a much smaller radius of \SI{0.84184 \pm 0.00067}{fm} from the spectroscopy method on muonic hydrogen~\cite{Pohl2010}, which posed a seven sigma discrepancy from the previous value. 

The PRad collaboration in 2019 measured the proton charge radius of \SI{0.831\pm0.014}{fm} using the elastic electron-proton scattering on a windowless gas target with a length of approximately \SI{40}{mm} along the beam direction~\cite{PRad2019}. Their result favors the smaller proton radius from the muonic hydrogen experiments. But their measurement also shows a proton electric form factor discrepancy beyond experimental uncertainty in the range of \SI{0.01}{}$\le Q^2 \le$\SI{0.06}{(GeV/\clight)^2} with the previous measurements~\cite{Bernauer2014}, introducing a new puzzle. 

Motivated by these two discrepancies, in this experiment, we used a newly developed cryogenic supersonic gas jet target to re-measure the proton electric form factors within the four-momentum transfer range of $0.01\le Q^2 \le $ \SI{0.045}{(GeV/\clight)^2}.  

\section{Experimental setup}
The experiment was performed in the A1 three-spectrometer facility at the Mainz Microtron (MAMI). MAMI can generate a high-quality unpolarized electron beam up to \SI{100}{\mu A} and energy up to \SI{1600}{MeV} with three stages of racetrack microtron and a fourth stage of double-sided microtron~\cite{Herminghaus:1976,Kaiser:2008,Dehn:2011}. We conducted this experiment at $I=$\SI{20}{\mu A} and $E_\text{beam} = $  \SI{315}{MeV} for the best beam stability and energy spread uncertainty.

In the A1 three-spectrometer facility, the scattered electrons can be detected by three high-resolution magnetic spectrometers, called A, B, and C. They can rotate horizontally around the interaction region, covering a wide range of scattering angles. Spectrometers A and C have a solid angle acceptance of up to $\Delta \Omega = $ \SI{28}{msr}, while spectrometer B can accept $\Delta \Omega = $ \SI{5.6}{msr}. In this experiment, we use spectrometer B for the actual cross-section measurement and spectrometer A as a luminosity monitor. 

The focal plane detector systems for all three spectrometers are very similar. They contain two scintillator planes, two vertical drift chambers (VDCs), and a gas Cherenkov detector. For this experiment, the scintillators are used for triggering and the VDCs  are used for the track reconstruction. The particle identification from the Cherenkov detector is not necessary for this measurement. The spectrometers have a relative momentum resolution of $\delta p/p = 10^{-4}$ and an angular resolution of $\delta \theta/\theta = $ \SI{3}{mrad} at the target. The detailed specifications for the spectrometers' magnets and detectors can be found in reference~\cite{Blomqvist:1998}.

This experiment uses a cryogenic supersonic jet target operated with molecular hydrogen. Unlike the existing extended windowless gas targets, it is designed to have a length of only about \SI{1}{mm} along the beam direction and to achieve an areal thickness of more than $\rho_\text{areal} = $ \SI[retain-unity-mantissa=false]{e18}{atoms/cm^2} when using hydrogen and running at a flow rate of \SI{2400}{l_n/h} with temperature $T_0 = $ \SI{40}{K}. It has a booster stage using liquid nitrogen for precooling and a cryogenic cold head for the second stage cooling. The cryogenic gas is then forced with high pressure to flow through a convergent-divergent nozzle into the target chamber under vacuum. The gas gets accelerated to supersonic velocities and adiabatically cools down during the expansion. Through rapid cooling, the gas forms clusters after the nozzle constriction. The jet then intersects with the electron beam at a right angle. It is then collected by a catcher several millimeters below. More details on the target can be found in reference~\cite{Grieser:2018qyq,A1Mainz2021}.

\begin{figure}
    \centering
    \includegraphics[width=0.38\textwidth]{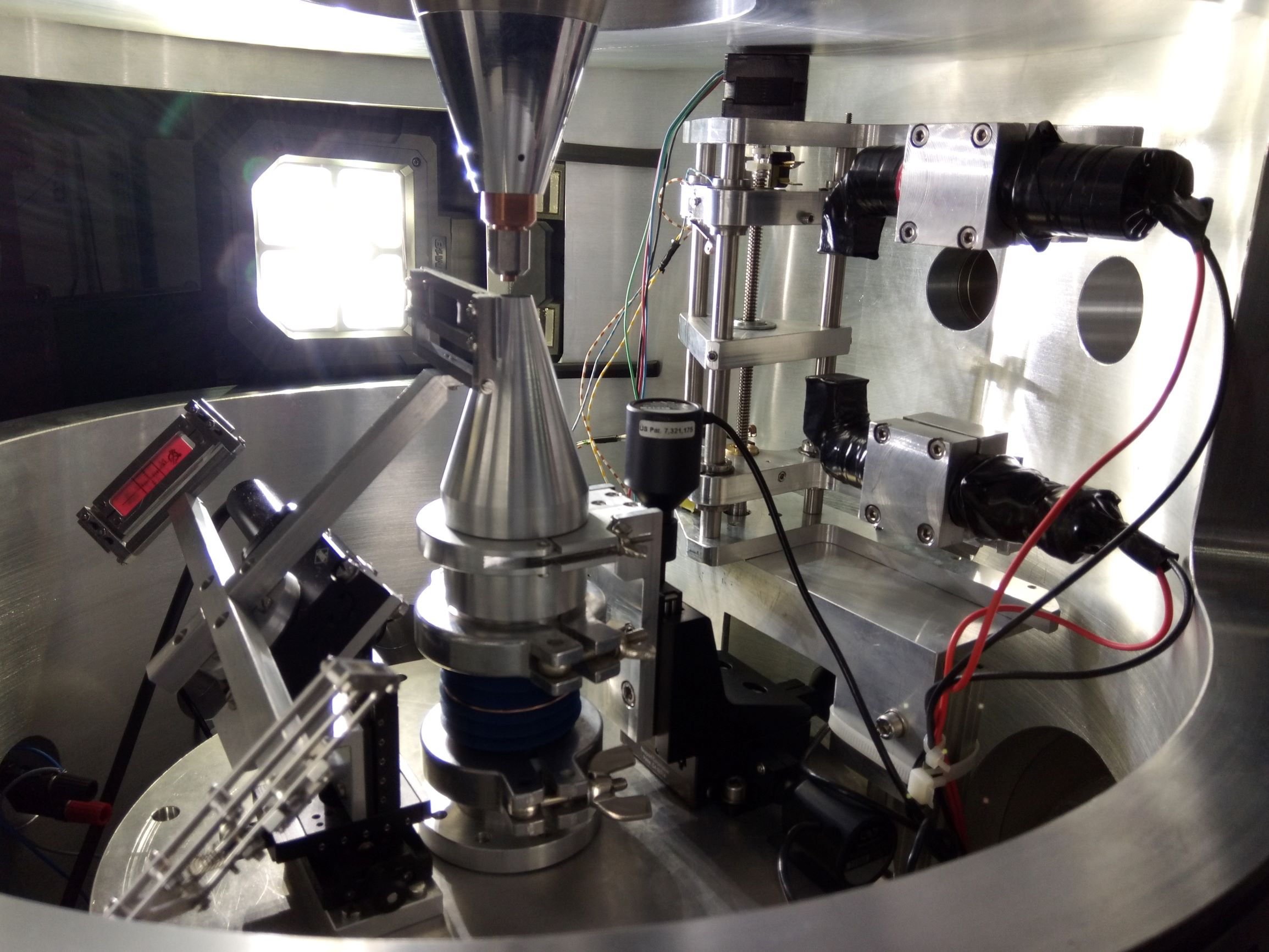}
    \caption{Photograph of the inside of the scattering chamber shows the target nozzle and catcher as well as the beam halo veto detector. The beam enters the chamber from the right and the beam halo collimators are placed further upstream~\cite{A1Mainz2021}.}
    \label{fig:target}
\end{figure}

To eliminate the background from the beam halo hitting the nozzle and catcher, a collimator and veto system is installed upstream of the target. The collimator consists of two vertically movable tungsten bricks located above and below the beam height. Each collimator brick has a thickness of $d = $ \SI{13}{cm} along the beam direction which can absorb electromagnetic showers from a primary electron energy of up to $E =$ \SI{1.5}{GeV}. The vertical position for each collimator brick can be individually optimized. The veto detector consists of two detector arms, made of scintillators and photomultiplier tubes, and can move vertically. It is located inside the scattering chamber right in front of the target to veto the residual beam halo electrons potentially hitting the catcher or the nozzle. The vertical position of the veto detectors can be individually optimized as well to best cover the nozzle and the catcher without intruding on the primary electron beam.

When designing this experiment, we prioritize cross-checking as many aspects as possible with the previous proton form factor measurement at the same facility back in 2010\cite{Bernauer2014}. Our gas-jet target system is fundamentally different from the old liquid hydrogen target. The beam halo collimator and veto aim to achieve zero background. Even though we still observe background events during the experiment, the new target design has a different and uncorrelated set of systematic uncertainties compared to the old target. Although we use the same spectrometers and focal plane detectors, we emphasize on the calibration of the spectrometer entrance collimator acceptance and the transfer matrix of the spectrometer magnetic field to minimize the chance of any possible error from these two sources.

\section{Theory and simulation}

\subsection{Elastic electron-proton scattering}

\begin{figure}
    \centering
    \includegraphics[width=0.3\textwidth]{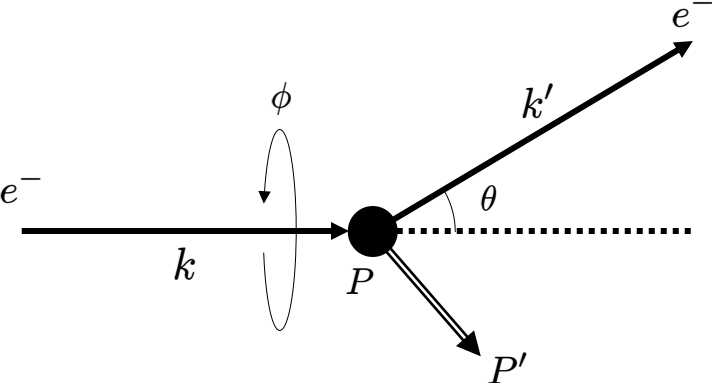}
    \caption{Kinematics of electron-proton scattering in laboratory frame.}
    \label{fig:ep_kinematics}
\end{figure}

The kinematics of electron-proton scattering in the laboratory frame is shown in Figure~\ref{fig:ep_kinematics}. The incoming electron has the four-momentum of $k = (E, \vec{p})$. The outgoing electron has the four-momentum of $k' = (E', \vec{p'})$ in the direction $(\theta, \phi)$. The virtual photon has the four-momentum $q=k-k'$. The target proton is initially at rest with four momentum $P=(m_p,\vec{0})$ and $P^\prime$ after scattering.

The electron is unpolarized, and the process is azimuthally symmetric. Therefore, it has only two degrees of freedom. One way to describe the process is to use the energy of the incoming electron $E$ and the scattering angle $\theta$. 
Since the virtual photon is space-like, we can define $Q^2$ as
\begin{equation}
    Q^{2}\equiv -q^{2}=4 E E^{\prime} \sin ^{2} \frac{\theta}{2},
\end{equation}
to be always positive. The polarization of the virtual photon is given by
\begin{equation}
    \varepsilon =\left(1+2(1+\tau) \tan ^{2} \frac{\theta}{2}\right)^{-1},
\end{equation}
where the dimensionless quantity $\tau = Q^2/(4m_p^2)$. $Q^2$ and $\varepsilon$ provide another way to specify this process.

In order to incorporate the internal structure of the proton, the Sachs electric and magnetic form factors $G_E$ and $G_M$ are introduced to describe the cross section 
\begin{multline}
    \left.\frac{\mathrm{d} \sigma}{\mathrm{d} \Omega}\right|_\text{lab} =\left(\frac{\mathrm{d} \sigma}{\mathrm{d} \Omega}\right)_{\mathrm{Mott}}\left[\frac{G_{E}^{2}\left(Q^{2}\right)+\tau G_{M}^{2}\left(Q^{2}\right)}{1+\tau} \right.\\
    \left. +2 \tau G_{M}^{2}\left(Q^{2}\right) \tan ^{2} \frac{\theta}{2}\right] 
    \\ = \left(\frac{\mathrm{d} \sigma}{\mathrm{d} \Omega}\right)_{\mathrm{Mott}} \frac{\varepsilon G_{E}^{2}\left(Q^{2}\right)+\tau G_{M}^{2}\left(Q^{2}\right)}{\varepsilon(1+\tau)},
\end{multline}
where the recoil-corrected Mott scattering cross section
\begin{equation}
    \left(\frac{d\sigma}{d\Omega}\right)_\text{Mott} = \frac{\alpha^{2}}{4 E^{2} \sin ^{4} \frac{\theta}{2}} \frac{E^{\prime}}{E}
\end{equation} 
is the scattering cross section of the electron on a point-like scalar particle.

At the limit of $Q^2=0$, the form factors are normalized to the electric charge and the magnetic moment. For the proton, $G_E(0)=1$ in the unit of electric charge and $G_M(0)=\mu_p$ in the unit of the nuclear magneton $\mu_N=(e\hbar)/(2m_p)$.

If we expand the proton electric form factor functions at low $Q^2$, we get
\begin{equation}
    G_E\left(Q^{2}\right) / G_E(0)=1-\frac{1}{6}\left\langle r^{2}\right\rangle Q^{2}+\frac{1}{120}\left\langle r^{4}\right\rangle Q^{4}-\ldots.
\end{equation}
We can then determine the proton charge radius as
\begin{equation}
   \left\langle r^{2}_{E}\right\rangle=-\left.\frac{6}{G_{E}(0)} \frac{\mathrm{d} G_{E}\left(Q^{2}\right)}{\mathrm{d} Q^{2}}\right|_{Q^{2}=0}.
   \label{eqn:proton_radius}
\end{equation}

\subsection{Radiative corrections}

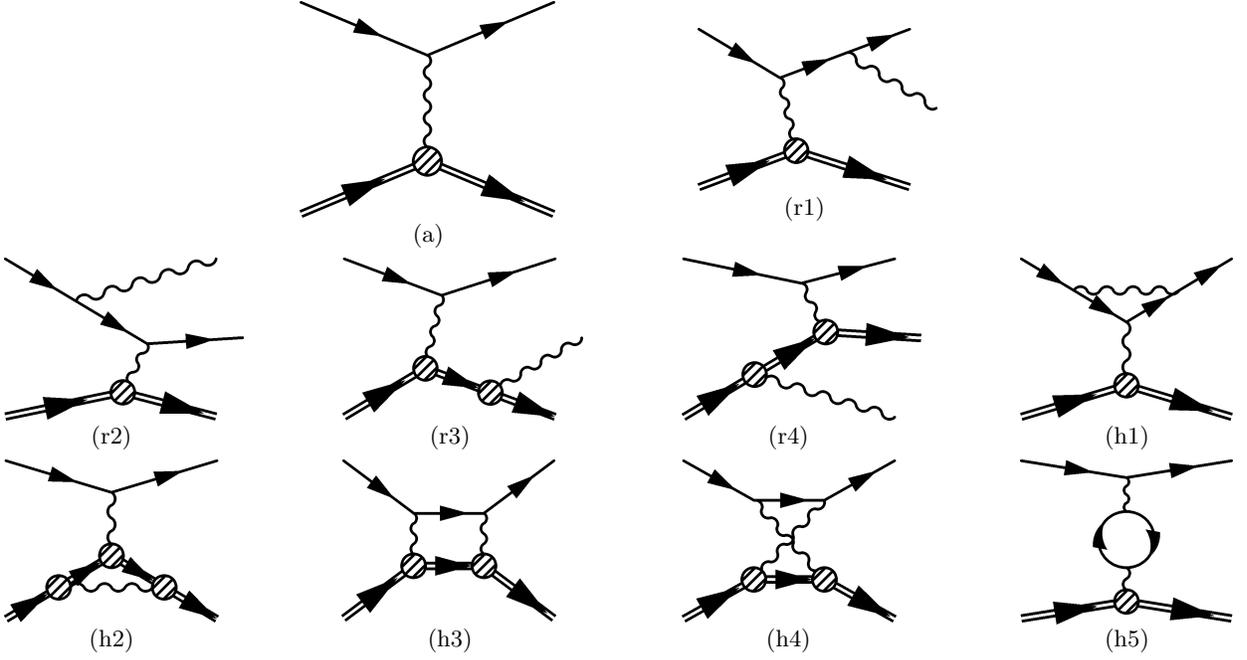
\begin{figure*}
    \centering
    \begin{fmffile}{diagram}
\begin{gather*}
\parbox{50mm}{
    \centering
    \begin{fmfgraph*}(120,80)
    \fmfleft{i1,i2}
    \fmfright{o1,o2}
    \fmf{heavy}{i1,v1,o1}
    \fmf{fermion}{i2,v2,o2}
    \fmf{photon,label=$\gamma$}{v1,v2}        \fmfblob{.09w}{v1}
    \fmfv{label=$e^-$}{i2}
    \fmfv{label=$e^-$}{o2}
    \fmfv{label=$p$}{i1}
    \fmfv{label=$p$}{o1}
    \end{fmfgraph*}\\
    (a)
} 
\parbox{50mm}{
    \centering
    \begin{fmfgraph}(100,60)
    \fmfleft{i1,i2}
    \fmfright{o1,p1,o2}
    \fmf{heavy}{i1,v1,o1}
    \fmfblob{.09w}{v1}
    \fmf{fermion}{i2,v2,v3,o2}
    \fmf{photon}{v1,v2}
    \fmffreeze
    \fmf{photon}{v3,p1}
    \end{fmfgraph}\\
    (r1)
}\\
\parbox{45mm}{
    \centering
    \begin{fmfgraph}(100,60)
    \fmfleft{i1,i2}
    \fmfright{o1,o2,p1}
    \fmf{heavy}{i1,v1,o1}
    \fmfblob{.09w}{v1}
    \fmf{fermion}{i2,v2,v3,o2}
    \fmf{photon}{v1,v3}
    \fmffreeze
    \fmf{photon}{p1,v2}
    \end{fmfgraph}\\
    (r2)
}
\parbox{45mm}{
    \centering
    \begin{fmfgraph}(100,60)
    \fmfleft{i1,i2}
    \fmfright{o1,p1,o2}
    \fmf{heavy}{i1,v1,v3,o1}
    \fmfblob{.09w}{v1,v3}
    \fmf{fermion}{i2,v2,o2}
    \fmf{photon}{v1,v2}
    \fmffreeze
    \fmf{photon}{v3,p1}
    \end{fmfgraph}\\
    (r3)
}
\parbox{45mm}{
    \centering
    \begin{fmfgraph}(100,60)
    \fmfleft{i1,i2}
    \fmfright{p1,o1,o2}
    \fmf{heavy}{i1,v1,v3,o1}
    \fmfblob{.09w}{v1,v3}
    \fmf{fermion}{i2,v2,o2}
    \fmf{photon}{v2,v3}
    \fmffreeze
    \fmf{photon}{p1,v1}
    \end{fmfgraph}\\
    (r4)
}
\parbox{45mm}{
    \centering
    \begin{fmfgraph}(100,60)
    \fmfleft{i2,i1}
    \fmfright{o2,o1}
    \fmf{heavy}{i2,v4,o2}
    \fmfblob{.09w}{v4}
    \fmf{fermion}{i1,v1,v2,v3,o1}
    \fmf{photon}{v4,v2}
    \fmffreeze
    \fmf{photon}{v1,v3}
    \end{fmfgraph}\\
    (h1)
}\\
\parbox{45mm}{
    \centering
    \begin{fmfgraph}(100,60)
    \fmfleft{i1,i2}
    \fmfright{o1,o2}
    \fmf{heavy}{i1,v1,v2,v3,o1}
    \fmfblob{.09w}{v1,v2,v3}
    \fmf{fermion}{i2,v4,o2}
    \fmf{photon}{v2,v4}
    \fmffreeze
    \fmf{photon}{v1,v3}
    \end{fmfgraph}\\
    (h2)
}
    \parbox{45mm}{
    \centering
    \begin{fmfgraph}(100,60)
    \fmfleft{i1,i2}
    \fmfright{o1,o2}
    \fmf{heavy}{i1,v1,v2,o1}
    \fmfblob{.09w}{v1}
    \fmfblob{.09w}{v2}
    \fmf{fermion}{i2,v3,v4,o2}
    \fmf{photon}{v1,v3}
    \fmf{photon}{v2,v4}
    \end{fmfgraph}\\
    (h3)
    }
     \parbox{45mm}{
     \centering
    \begin{fmfgraph}(100,60)
    \fmfleft{i1,i2}
    \fmfright{o1,o2}
    \fmf{heavy, tension=2}{i1,v1}
    \fmf{heavy}{v1,v2}
    \fmf{heavy, tension=2}{v2,o1}
    \fmfblob{.09w}{v1}
    \fmfblob{.09w}{v2}
    \fmf{fermion, tension=2}{i2,v3}
    \fmf{fermion}{v3,v4}
    \fmf{fermion, tension=2}{v4,o2}
    \fmf{photon}{v1,v4}
    \fmf{photon}{v2,v3}
    \end{fmfgraph}\\
    (h4)
    }
     \parbox{45mm}{
     \centering
    \begin{fmfgraph}(100,60)
    \fmfleft{i1,i2}
    \fmfright{o1,o2}
    \fmf{heavy}{i1,v1,o1}
    \fmfblob{.09w}{v1}
    \fmf{fermion}{i2,v2,o2}
    \fmf{photon}{v3,v2}
    \fmf{photon}{v4,v1}
    \fmf{fermion,left,tension=.3}{v4,v3,v4}
    \end{fmfgraph}   \\
    (h5)
    }
\end{gather*}
\end{fmffile}
    \caption[Feynman diagrams for electron-proton scattering and radiative corrections]{Feynman diagrams for electron-proton scattering and radiative corrections: (a) is the leading order diagram for elastic electron-proton scattering; (r) are the leading order diagrams for bremsstrahlung with (r1)(r2) on the electron and (r3)(r4) on the proton; (h) are the next to leading order for elastic electron-proton scattering. (h1) is the electron vertex correction. (h2) is the proton vertex correction. (h3) and (h4) are the box and crossed-box diagram and (h5) is the vacuum polarization. }
    \label{fig:rad}
\end{figure*}

The Feynman diagram for the leading term for the unpolarized electron-proton scattering is shown as diagram (a) in Figure~\ref{fig:rad}. The cross section from this leading term is proportional to the square of the fine structure constant $\alpha$. The emission of the soft bremsstrahlung photons, as shown in diagrams (r) in Figure~\ref{fig:rad}, goes undetected. The photons remove a portion of the energy from the proton or the electron, changing the kinematic parameters of the scattering and the final electron momenta we measure in the spectrometer. Therefore, we should see events with electrons falling in a long tail towards lower momentum in our measurements. The effect should be taken into account in the cross section measurement for the elastic electron-proton scattering.

The bremsstrahlung diagrams are of order $\alpha^3$ and diverge at the infrared limit for the photon. One can get around by introducing higher-order terms of the elastic scattering, as shown in diagrams (h) in Figure~\ref{fig:rad}. The interference terms of the leading order and higher-order for the elastic scattering have divergences that cancel out those in the bremsstrahlung terms. However, diagrams (h1)(h2)(h5) in Figure~\ref{fig:rad} are logarithmically divergent for large momenta. This divergence can be fixed by using the BPHZ renormalization method as shown in \cite{Vanderhaeghen2000}.

All these above introduces a correction to the measured cross section:
\begin{equation}
\left(\frac{\mathrm{d} \sigma}{\mathrm{d} \Omega}\right)_\text{exp}=\left(\frac{\mathrm{d} \sigma}{\mathrm{d} \Omega}\right)_{0}
\left(1+\delta\right).
\end{equation}
A proper evaluation of $\delta$ is essential to retrieve the elastic scattering cross section from what we measure in the experiment. Several terms contribute at the order of $\alpha^3$ but with divergence by themselves (we use $\mathcal{M}$ to denote scattering amplitude):
\begin{enumerate}
    \item Bremsstrahlung on electron: $|\mathcal{M}_{r1}+\mathcal{M}_{r2}|^2$\\
    The divergence in this term is canceled out by $2\text{Re}[\mathcal{M}_{a}^\dagger \mathcal{M}_{h1}]$.
    
    \item Bremsstrahlung on proton:$|\mathcal{M}_{r3}+\mathcal{M}_{r4}|^2$\\
    The divergence in this term is canceled out by $2\text{Re}[\mathcal{M}_{a}^\dagger \mathcal{M}_{h2}]$.
    
    \item Interference term: $2\text{Re}[(\mathcal{M}_{r1}+\mathcal{M}_{r2})]^\dagger(\mathcal{M}_{r3}+\mathcal{M}_{r4})$\\
    The divergence in this term is canceled out by $2\text{Re}[\mathcal{M}_{a}^\dagger (\mathcal{M}_{h3}+\mathcal{M}_{h4})]$.
\end{enumerate}

The only remaining term of order $\alpha^3$ is $2\text{Re}[\mathcal{M}_{a}^\dagger \mathcal{M}_{h5}]$ which does not contain any divergence. 

The following lists all the correction terms used in this experiment after the IR divergence is canceled out. For detailed calculations, please refer to \cite{Vanderhaeghen2000} and \cite{Maximon2000}.

\begin{itemize}
    \item The vacuum polarization, $2\text{Re}[\mathcal{M}_{a}^\dagger \mathcal{M}_{h5}]$
        \begin{equation}
        \delta_\text{vac} =\frac{\alpha}{\pi} \frac{2}{3}\left\{\left(v^{2}-\frac{8}{3}\right)+v \frac{3-v^{2}}{2} \ln \left(\frac{v+1}{v-1}\right)\right\} 
        \end{equation}
        where $v$ is given by 
        \begin{equation}
        v^{2} \equiv 1+\frac{4 m_{l}^{2}}{Q^{2}},
        \end{equation}
        and $m_{l}$ is the mass of the lepton in the loop. 
    \item The bremsstrahlung on the electron, $|\mathcal{M}_{r1}+\mathcal{M}_{r2}|^2$
        \begin{equation}
        \begin{aligned} \delta_{R}=& \frac{\alpha}{\pi}\left\{\ln \left(\frac{\left(\Delta E_{s}\right)^{2}}{E \cdot E_{e l}^{\prime}}\right)\left[\left(\frac{Q^{2}}{m^{2}}\right)-1\right]-\frac{1}{2} \ln ^{2} \eta \right.\\ &\left.+\frac{1}{2} \ln ^{2}\left(\frac{Q^{2}}{m^{2}}\right)-\frac{\pi^{2}}{3}+\operatorname{Sp}\left(\cos ^{2} \frac{\theta_{e}}{2}\right)\right\} ,
        \end{aligned}
        \label{eqn:brem_electron}
        \end{equation}
        where the Spence function is defined by
        \begin{equation}
        \text{Sp}(x) \equiv-\int_{0}^{x} d t \frac{\ln (1-t)}{t},
        \end{equation}
        and 
        \begin{equation}
                    \eta=E / E_{el}^{\prime}, \quad \Delta E_{s}=\eta\left(E_{el}^{\prime}-E_{e}^{\prime}\right),
        \end{equation}
        in which $E_{el}^{\prime}$ is the energy of the elastically scattered electron without emitting photon and $E_{e}^{\prime}$ is the energy of eletron after photon emission. The difference between these two quantities $\Delta E^{\prime}$ is called the cut-off energy, which is typically determined by the detector's acceptance. The correction term covers all photon emissions with energy up to the cut-off energy.
    \item The vertex correction term on the electron side  $2\text{Re}[\mathcal{M}_{a}^\dagger \mathcal{M}_{h1}]$
    \begin{equation}
    \delta_\text{vertex}=\frac{\alpha}{\pi}\left\{\frac{3}{2} \ln \left(\frac{Q^{2}}{m^{2}}\right)-2-\frac{1}{2} \ln ^{2}\left(\frac{Q^{2}}{m^{2}}\right)+\frac{\pi^{2}}{6}\right\},
    \end{equation}
    and the divergence is canceled by the one in the bremsstrahlung on the electron.

    \item The vertex correction term on the proton side $|\mathcal{M}_{r3}+\mathcal{M}_{r4}|^2$.
    This is more complicated as it involves the internal structure of the proton.
    Maximon and Tjon\cite{Maximon2000} divided this into three parts: one proportional to $Z$(hadron charge): $\delta_1$, one proportional to $Z^2$: $\delta_2$, and one from the inclusion of the form factor for the proton: $\delta_{el}^{(1)}$. 
    \begin{multline}
            \delta_{1}=  \frac{2 \alpha}{\pi}\left\{\ln \left(\frac{4\left(\Delta E_{s}\right)^{2}}{Q^{2} x}\right) \ln \eta \right.\\ 
            \left.+\operatorname{Sp}\left(1-\frac{\eta}{x}\right)-\operatorname{Sp}\left(1-\frac{1}{\eta x}\right)\right\}
        \label{eqn:brem_proton1}
    \end{multline}
    \begin{multline}
            \delta_{2}= \frac{\alpha}{\pi}\left\{\ln \left(\frac{4\left(\Delta E_{s}\right)^{2}}{m_{p}^{2}}\right)\left(\frac{E_{p}^{\prime}}{\left|\vec{p^{\prime} _p}\right|} \ln x-1\right)+1\right.\\ 
            +\frac{E_{p}^{\prime}}{\left| \vec{p^{\prime}_{p}}\right|}
    \left(-\frac{1}{2} \ln ^{2} x-\ln x \ln \left(\frac{\rho^{2}}{m_{p}^{2}}\right)+\ln x\right.\\ \left.\left.-\operatorname{Sp}\left(1-\frac{1}{x^{2}}\right)+2 \operatorname{Sp}\left(-\frac{1}{x}\right)+\frac{\pi^{2}}{6}\right)\right\},
    \label{eqn:brem_proton2}
    \end{multline}

    where 
    \begin{equation}
    x=\frac{(Q+\rho)^{2}}{4 m_{p}^{2}}, \quad \rho^{2}=Q^{2}+4 m_{p}^{2}.
    \end{equation}
    Maximon and Tjon found the term $\delta_{el}^{(1)}$ to be much smaller than the other contributions in $\delta_2$ in the range of energies and momentum transfers for this experiment\cite{Maximon2000}.
\end{itemize}

Higher order radiative corrections can be approximated by exponentiating the first order vertex and real radiative corrections \cite{Bloch1937,Yennie1961}. Vanderhaeghen \emph{et al.} showed the vacuum polarization, by iterating the first order to all orders, introduces a non-exponentiation term\cite{Vanderhaeghen2000}, leaving us the final result:

\begin{equation}
\left(\frac{\mathrm{d} \sigma}{\mathrm{d} \Omega}\right)_\text{exp}=\left(\frac{\mathrm{d} \sigma}{\mathrm{d} \Omega}\right)_{0} \frac{e^{\delta_\text{vertex}+\delta_{R}+ \delta_{1}+ \delta_{2}}}{\left(1-\delta_\text{vac} / 2\right)^{2}}.
\end{equation}

In practice, we use the exponential form
\begin{equation}
\left(\frac{\mathrm{d} \sigma}{\mathrm{d} \Omega}\right)_\text{exp}\left(\Delta E^{\prime}\right)=\left(\frac{\mathrm{d} \sigma}{\mathrm{d} \Omega}\right)_{0} e^{\delta_\text{vertex}+\left[\delta_{R}+\delta_{1}+\delta_{2}\right]\left(\Delta E^{\prime}\right)},
\end{equation}
and it introduced an error below 0.05\% for the kinematics covered in this experiment.

\subsection{Simulation of the primary process}

The primary process generator for this experiment generates the out-going electrons to be detected from the electron-proton scattering process, calculates the cross section, and applies the radiative correction. The radiative process generator used in this experiment is adapted from the OLYMPUS experiment\cite{schmidt2017} which is based on the previous version of the radiative generator developed at A1@MAMI\cite{Bernauer2010thesis}. 

Although using the same set of corrections, our new radiative generator is re-developed independently from the old software package to avoid any software mistakes. The output of our generator is also compared against the non-exponentiated approach in the OLYMPUS generator\cite{schmidt2017} and the ESEPP generator\cite{Gramolin2014} for a better understanding of the systematic uncertainties.

The generator follows several steps to generate all the information we need for a single event:
\begin{enumerate}

    \item Generate the elastically scattered electron's direction and momentum
   
    \item Generate the electron energy loss $\Delta E^{\prime}$
    
    \item Generate the photon direction relative to the electron $\theta_{e\gamma}$
    
    \item Correct the cross section based on $\Delta E^{\prime}$ and $\theta_{e\gamma}$
\end{enumerate}

In the first step, for each event, we generate the scattering angle and the azimuthal angle pseudo-randomly within the vicinity of the angular acceptance of the spectrometer. The available angular phase space is slightly wider because of the finite distribution of the primary vertex position from the gas jet target. The final acceptance or rejection for each simulation event is based on both the vertex position and the scattered electron direction. 

\subsubsection{Radiative tail $\Delta E^{\prime}$}

The correction from the electron contribution, as mentioned in Equation~\ref{eqn:brem_electron} has the $\Delta E^{\prime}$ dependent term as 
\begin{equation}
    \delta_{r1r2}(\Delta E^{\prime})= \frac{\alpha}{\pi}\ln \left(\frac{\left(\Delta E_{s}\right)^{2}}{E \cdot E_{e l}^{\prime}}\right)\left[\left(\frac{Q^{2}}{m^{2}}\right)-1\right].
\end{equation}
Recall that $E$ is the incoming electron energy and $E_{el}^{\prime}$ is the energy of the electron assuming scattered elastically at the specified angle. $Q^2$ is the momentum transfer. $\Delta E_s$ is the energy of the radiated photon in the center of mass frame of the photon and the proton. Replacing $\eta = E/ E_{el}^{\prime}$ and $\Delta E_s = \eta \Delta E^{\prime}$, we have in the exponential form:
\begin{equation}
        \text{exp}(\delta_{r1r2})=\left(\frac{\left(\Delta E_{s}\right)^{2}}{E \cdot E_{e l}^{\prime}}\right)^a
        = \eta^a \left(\frac{\Delta E^{\prime}}{E^{\prime}_{el}}\right)^{2a},
\end{equation}
in which 
\begin{equation}
    a =  \frac{\alpha}{\pi}\left[\left(\frac{Q^{2}}{m^{2}}\right)-1\right].
\end{equation}

Similarly, there are two terms on the proton side contributing to the radiative tail that depend on $\Delta E^\prime$. From Equation~\ref{eqn:brem_proton1} for $\delta_{1}$, we have
\begin{equation}
    \delta_{1}(\Delta E^{\prime})= \frac{2 \alpha}{\pi}\left[\ln \left(\frac{4\left(\Delta E_{s}\right)^{2}}{Q^{2} x}\right) \ln \eta\right],
\end{equation}    
and its exponential form can be written in the same way
\begin{equation}
    \text{exp}(\delta_{1})= \left(\frac{4\left(\Delta E_{s}\right)^{2}}{Q^{2} x}\right)^b = \left(\frac{4 E^2}{Q^{2} x}\right)^b \left(\frac{\Delta E^{\prime}}{E^{\prime}_{el}}\right)^{2b}
\end{equation}
where 
\begin{equation}
    b= \frac{2 \alpha}{\pi}\ln \eta,
\end{equation}

The other term $\delta_{2}$ from Equation~\ref{eqn:brem_proton2} gives
\begin{equation}
    \delta_{2}(\Delta E^{\prime})= \frac{\alpha}{\pi}\left[\ln \left(\frac{4\left(\Delta E_{s}\right)^{2}}{m_{p}^{2}}\right)\left(\frac{E_{P}^{\prime}}{\left|\vec{p}^{\prime}_P\right|} \ln x-1\right)\right]
\end{equation}
and the exponential form 
\begin{equation}
    \text{exp}(\delta_{2})= \left(\frac{4 E^2}{m_{p}^{2}}\right)^c \left(\frac{\Delta E^{\prime}}{E^{\prime}_{el}}\right)^{2c}
\end{equation}
where 
\begin{equation}
    c= \frac{2 \alpha}{\pi}\left(\frac{E_{P}^{\prime}}{\left|\vec{p}^{\prime}_P\right|} \ln x-1\right).
\end{equation}

Combining all these, we have the following expression describing the radiative tail:
\begin{equation}
    \text{exp}(\delta_{r1r2}+\delta_{1}+\delta_{2})
    = \underbrace{\eta^a \left(\frac{4 E^2}{Q^{2} x}\right)^b \left(\frac{4 E^2}{m_{p}^{2}}\right)^c}_{\text{Weight term}} \underbrace{\left(\frac{\Delta E^{\prime}}{E^{\prime}_{el}}\right)^{t}}_{\text{Distribution term}}
    \label{eqn:rad_weight}
\end{equation}
where the last term $\left(\frac{\Delta E^{\prime}}{E^{\prime}_{el}}\right)^{t}$ includes all $\Delta E^\prime$ dependence and $t=2a+2b+2c$.

From Equation~\ref{eqn:rad_weight}, we need to sample according to the distribution term and assign the weight term to produce the correct shape of the radiative tail. The distribution of $\Delta E^{\prime}$ should follow:

\begin{equation}
    \int_0^{\Delta E^{\prime}}P_{E^{\prime}_{el},t}(\epsilon) \mathrm{d}\epsilon = \left(\frac{\Delta E^{\prime}}{E^{\prime}_{el}}\right)^{t}.
    \label{eqn:tail_cdf}
\end{equation}

One can easily verify the distribution is normalized to 1 by taking the cut-off energy, $\Delta E^{\prime}$, to the maximum limit of $E^{\prime}_{el}$ (ignore the electron mass). The exact expression for the distribution can be calculated by taking the derivative of Equation~\ref{eqn:tail_cdf}:

\begin{equation}
    P_{E^{\prime}_{el},t}(\Delta E^{\prime}) = \frac{t}{\Delta E^{\prime}}\left(\frac{\Delta E^{\prime}}{E^{\prime}_{el}}\right)^{t}.
\end{equation}

To generate events that follow this distribution, we can use the inverse transform technique by setting:
\begin{equation}
    \Delta E^{\prime} = E_{el}^{\prime} \cdot u^{\frac{1}{t}},
\end{equation} 
where $u$ is uniformly sampled in $[0,1]$. Then we can calculate and assign the weight factor, according to Equation~\ref{eqn:rad_weight}, to this event to generate the correct radiative tail.

\subsubsection{Photon direction $\theta_{e\gamma}$}

The cross section for photon emission has high variance even locally around the electron directions, where the cross section varies over many orders of magnitude in a small range of relative angles. 

Therefore, we use importance sampling to address this issue when generating the photon direction. To do so, we instead use a probability distribution that has two properties. First, this distribution needs to be close to the actual distribution, so the weight for each generated event is more uniform. Second, this distribution needs to be numerically easy to sample from.

The distribution we used to model the photon direction is 
\begin{equation}
    \begin{aligned} P(\cos \theta_{e\gamma}) &=\frac{1}{2 \frac{E}{|\vec{p}|} \log \left[\frac{E+|\vec{p}|}{E-|\vec{p}|}\right]-4} \times \frac{1-\cos ^{2} \theta_{e\gamma}}{\left(\frac{E}{|\vec{p}|}-\cos \theta_{e\gamma}\right)^{2}} \\ &=\frac{1}{N} \times \frac{1-\cos ^{2} \theta_{e\gamma}}{\left(\frac{E}{|\vec{p}|}-\cos \theta_{e\gamma}\right)^{2}} \end{aligned}
\end{equation}
and the cumulative distribution is given by
\begin{equation}
\begin{aligned}
    F(\cos \theta_{e\gamma})=&\frac{1}{N} \int_{-1}^{1} d \cos \theta_{e\gamma} \frac{1-\cos ^{2} \theta_{e\gamma}}{\left(\frac{E}{|\vec{p}|}-\cos \theta_{e\gamma}\right)^{2}}\\
    =&\frac{1}{N}\left[\cos \theta_{e\gamma}+2 \frac{E}{|\vec{p}|} \log \left\{\frac{\frac{E}{|\vec{p}|}-\cos \theta_{e\gamma}}{\frac{E}{|\vec{p}|}-1}\right\}\right.\\
    &\left.-2-\frac{E}{|\vec{p}|}-\frac{\left(1-\frac{E}{|\vec{p}|}^{2}\right)}{\frac{E}{|\vec{p}|}-\cos \theta_{e\gamma}}\right].
\end{aligned}
\end{equation}
Similarly to before, we can use a uniform distribution generator with the inversion of $F(\cos \theta_{e\gamma})$ to generate such a distribution. The inversion of $F$ is numerically evaluated using the bisection method. 

For each event, we randomly pick the incoming or outgoing electron with equal probability. Then we sample $\theta_{e\gamma}$ as described here and $\phi_{e\gamma}$ isotropically. Finally, we have all the required kinematic variables, $\Omega_e, \Omega_\gamma, \Delta E^\prime$ to specify and calculate all the weights for this event. 

\subsection{Simulation of other aspects}

Besides generating the primary scattering process, the simulation also simulates part of the target system and the detectors to evaluate the acceptance.

\subsubsection{Target system}
We simulate the target by generating the primary vertex of the scattering with a probability distribution from the sum of two Gaussian distributions with different standard deviations to mimic the observed distribution. 

The vertex distribution we observed in the data has two contributions. The central jet and the ambient gas in the chamber interacting with the electron beam produces a heavy-tail distribution in the $z$ direction independent of the spectrometer angles. When we reconstruct the projected vertex $z$ position along the beamline from the position $y_0$ perpendicular to the central plane of the spectrometer and the in-plane angle $\phi$, their finite resolution creates a vertex distribution that has the spectrometer angle dependence. Therefore, we obtain the standard deviation of the central gas jet distribution and the vertex resolution from a fit of the angular dependence of the reconstructed vertex $z$ distribution from the data. We combine it with an extensive Gaussian distribution to simulate the ambient hydrogen gas in the target chamber. The fifth and sixth plot in Figure~\ref{fig:data_vs_simul} shows the good agreement between this model of target and the data for both the central jet and the ambient gas in the chamber. 

We also simulate the energy loss and multiple scattering of the beam and scattered electron in the target. Since the gas jet has very limited density, these two effects have an impact on the order of \SI{10}{keV} compared to the electron energy, which is close to \SI{300}{MeV}. This impact is much smaller than the other uncertainties. 

The simulation also includes the energy loss for scattered electrons leaving the target chamber and entering the spectrometers. There are two vacuum windows (thin Kapton foils) because the vacuum of the chamber and the detector is maintained separately. Both walls have a small material thickness, so this is simulated by direct calculation using the empirical equations of the energy loss~\cite{PDG} and multiple scattering in the wall material.

\subsubsection{Detector acceptance and resolution}
\begin{figure}
    \centering
    \includegraphics[width=0.22\textwidth]{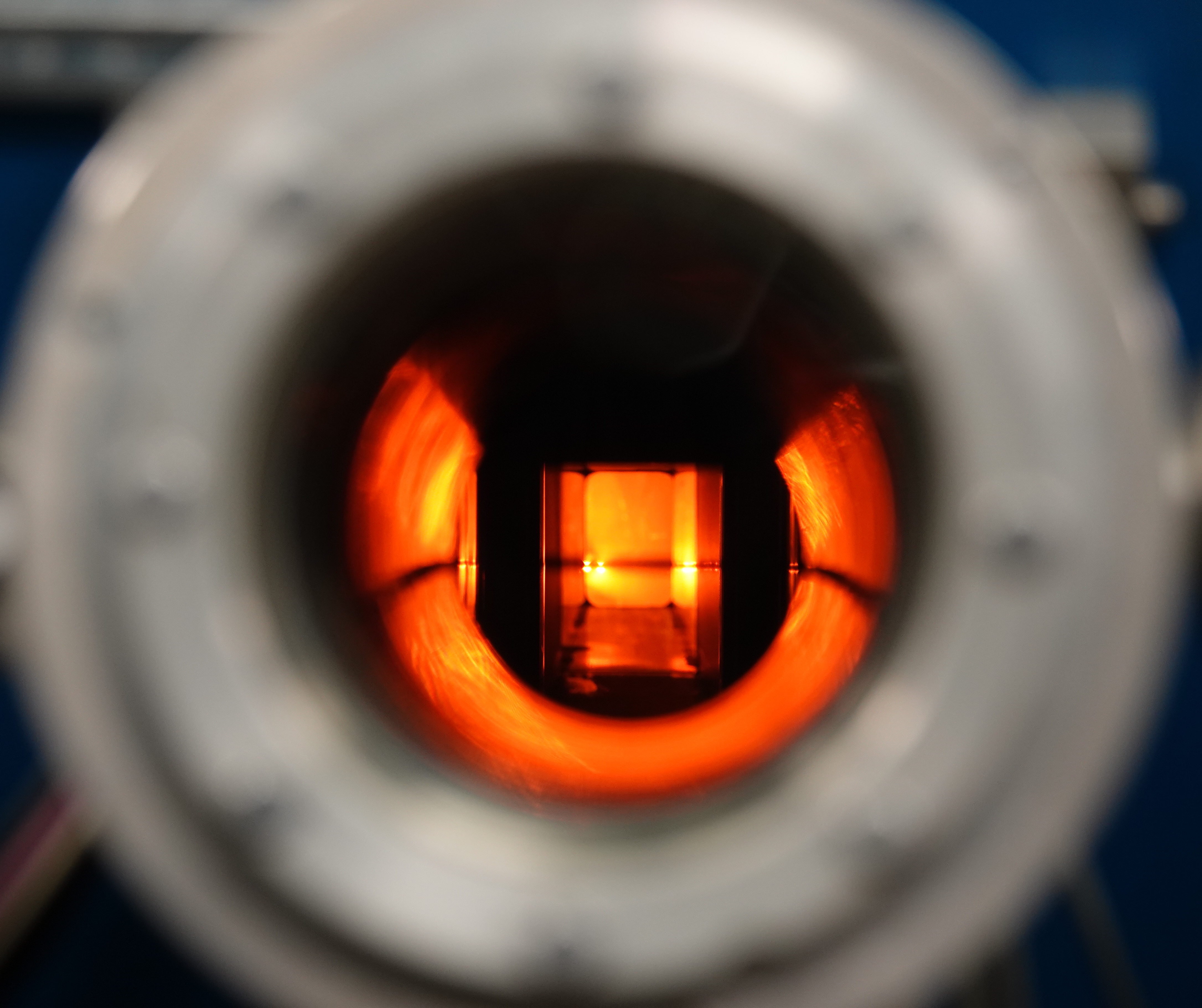}
    \includegraphics[width=0.22\textwidth]{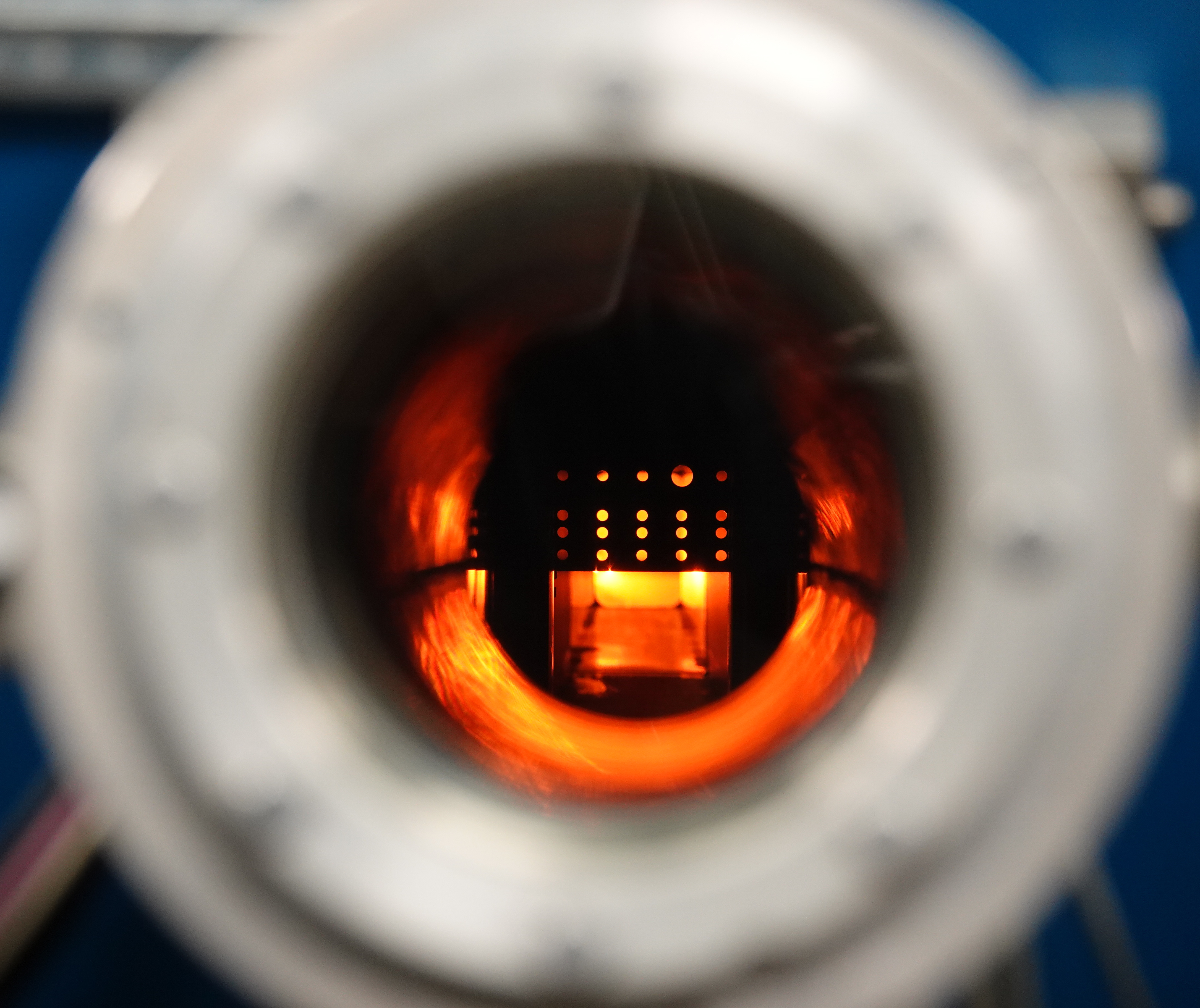}
    \caption[View through the window on the back of spectrometer B]{View through the window on the back of spectrometer B. Left: Sieve slit fully retracted with visual for the vacuum window. Right: Sieve slit half-in. During the calibration process of the transfer matrix parameters, the sieve slit is fully extended, allowing electrons only from a known set of target coordinates to pass.}
    \label{fig:spec_window}
\end{figure}

We do not include the full simulation of the focal plane detectors. Instead, we only propagate the particle as far as the entrance of the spectrometer to simulate the angular acceptance of the spectrometer. We model the effects from track reconstructions in the VDC and the back-propagation using the transfer matrix by quantifying the momentum and angular resolutions directly.

The angular acceptance is governed by the entrance collimator of the spectrometers. This acceptance collimator has a well-defined geometry. The location of the collimator edges with respect to the entrance of the spectrometer is calibrated by the theodolite light measurement. Each of the spectrometers has a glass window behind the magnet that allows a direct view of the acceptance collimator and the target chamber, as shown in Figure~\ref{fig:spec_window}. With proper lighting and the sieve slit collimator in position, one can look into the window through a theodolite and measure the relative angle of sight for each of the holes on the sieve slit and the acceptance collimator edges. With the measurement results, the known size and locations of the sieve slit holes, the sieve slit position, and the acceptance collimator position from the center of the target, we can calculate the exact location of the entrance collimator edges and determine the precise angular acceptance relative to the central angle of the spectrometer.

The angular and momentum resolution of the detectors are known to approximate values. The exact value is deduced from the data itself as the condition of the detectors might change very slightly from experiment to experiment. This improved estimation of the resolution helps reduce the systematic uncertainty of from the choice cut-off energy for the radiative tail by approximately 5\%. We use three parameters to model each of the angular and momentum resolutions: one parameter for the standard resolution, one parameter for the poorer resolution when the VDC reconstruction is sub-optimal, and one ratio parameter to quantify how often this sub-optimal reconstruction happens.

\begin{figure*}
    \centering
    \includegraphics[width=0.32\textwidth]{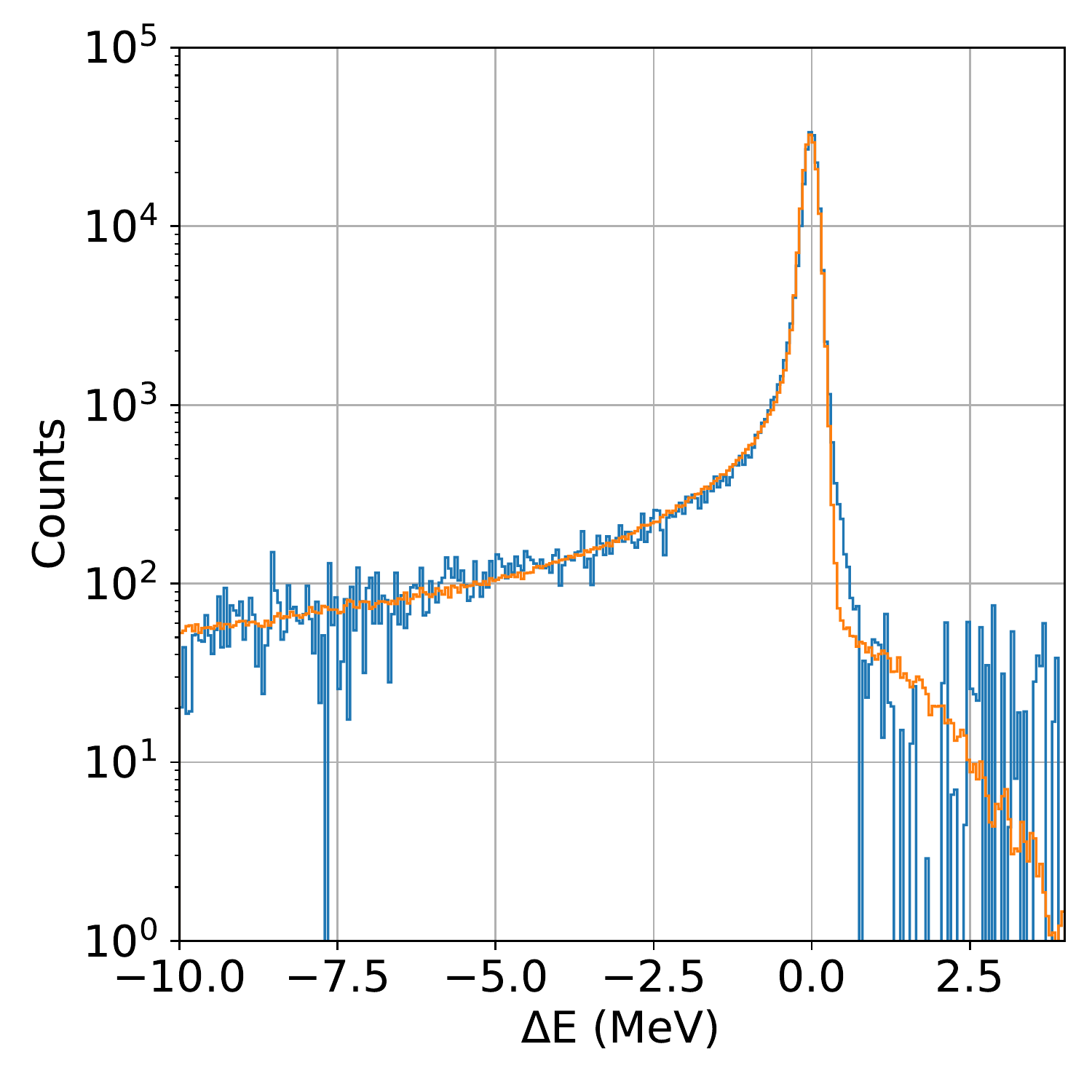}
    \includegraphics[width=0.31\textwidth]{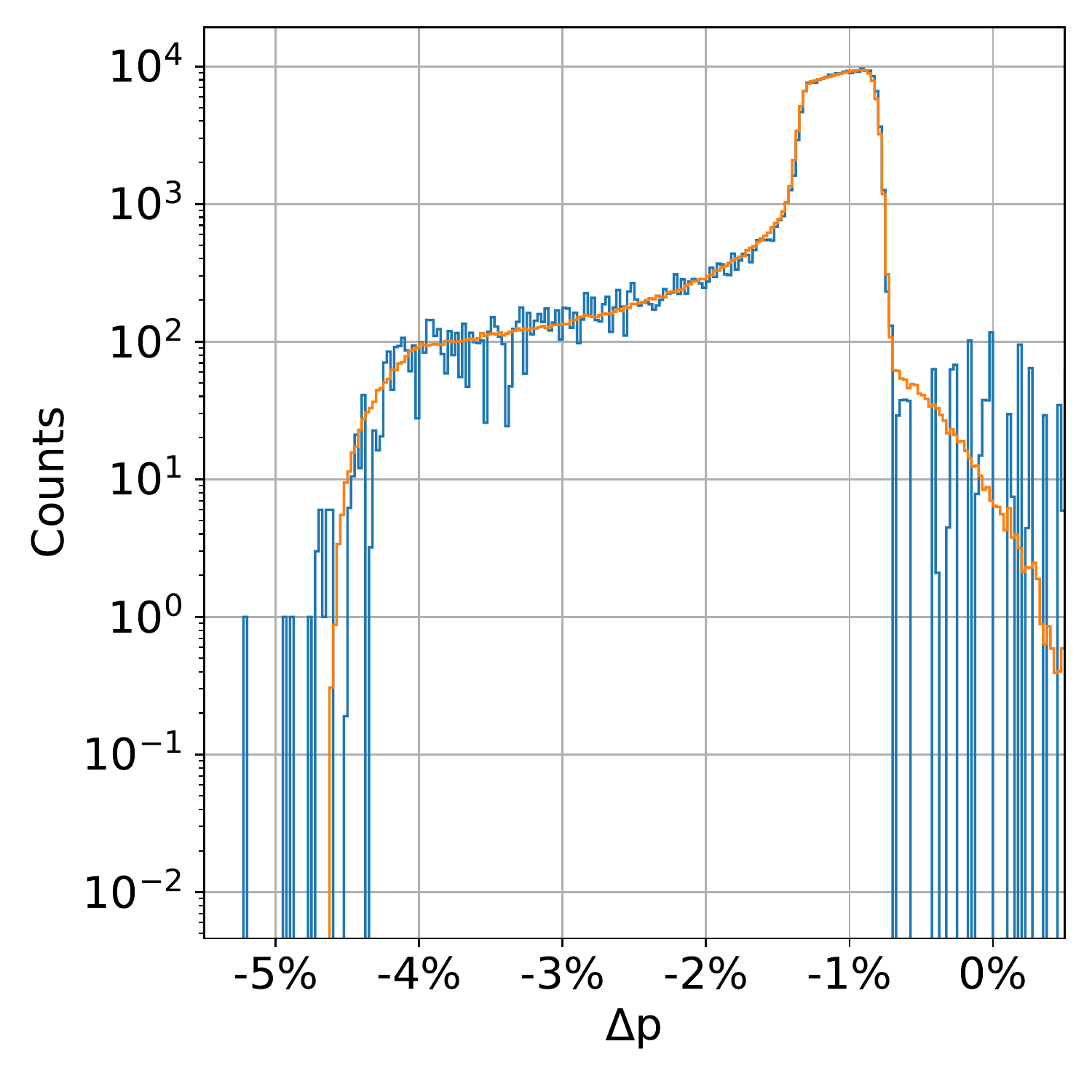}
    \includegraphics[width=0.32\textwidth]{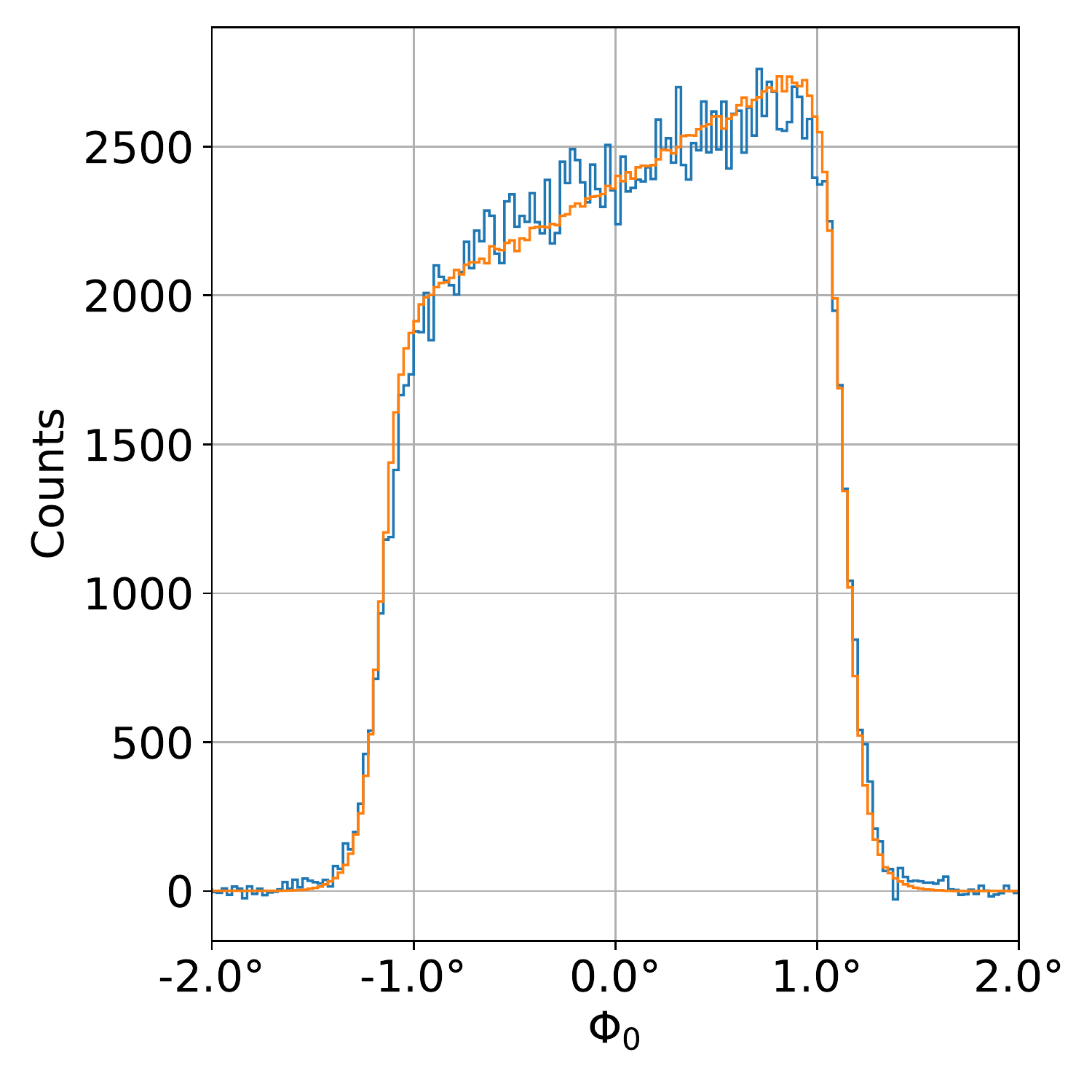}
    \includegraphics[width=0.32\textwidth]{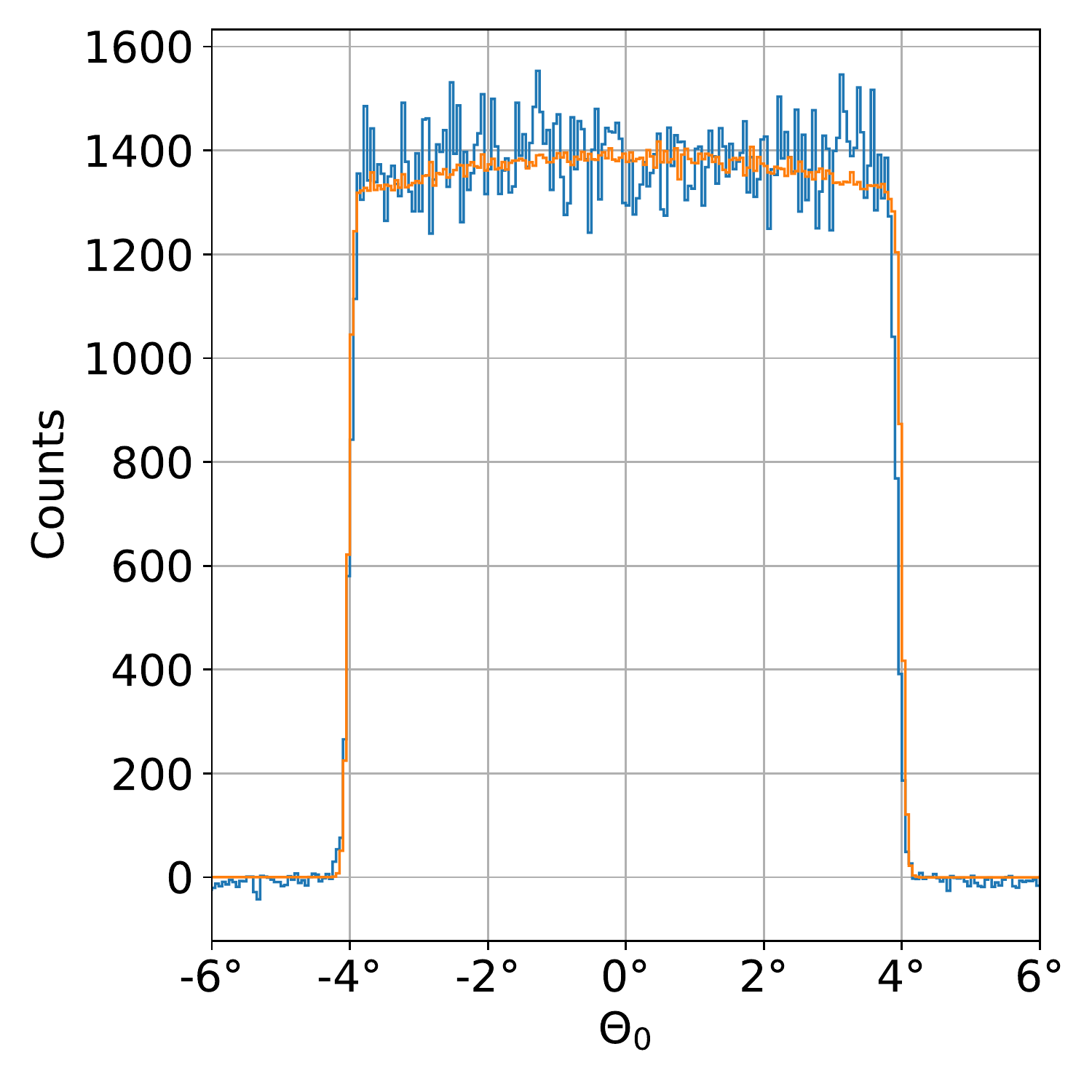}
    \includegraphics[width=0.32\textwidth]{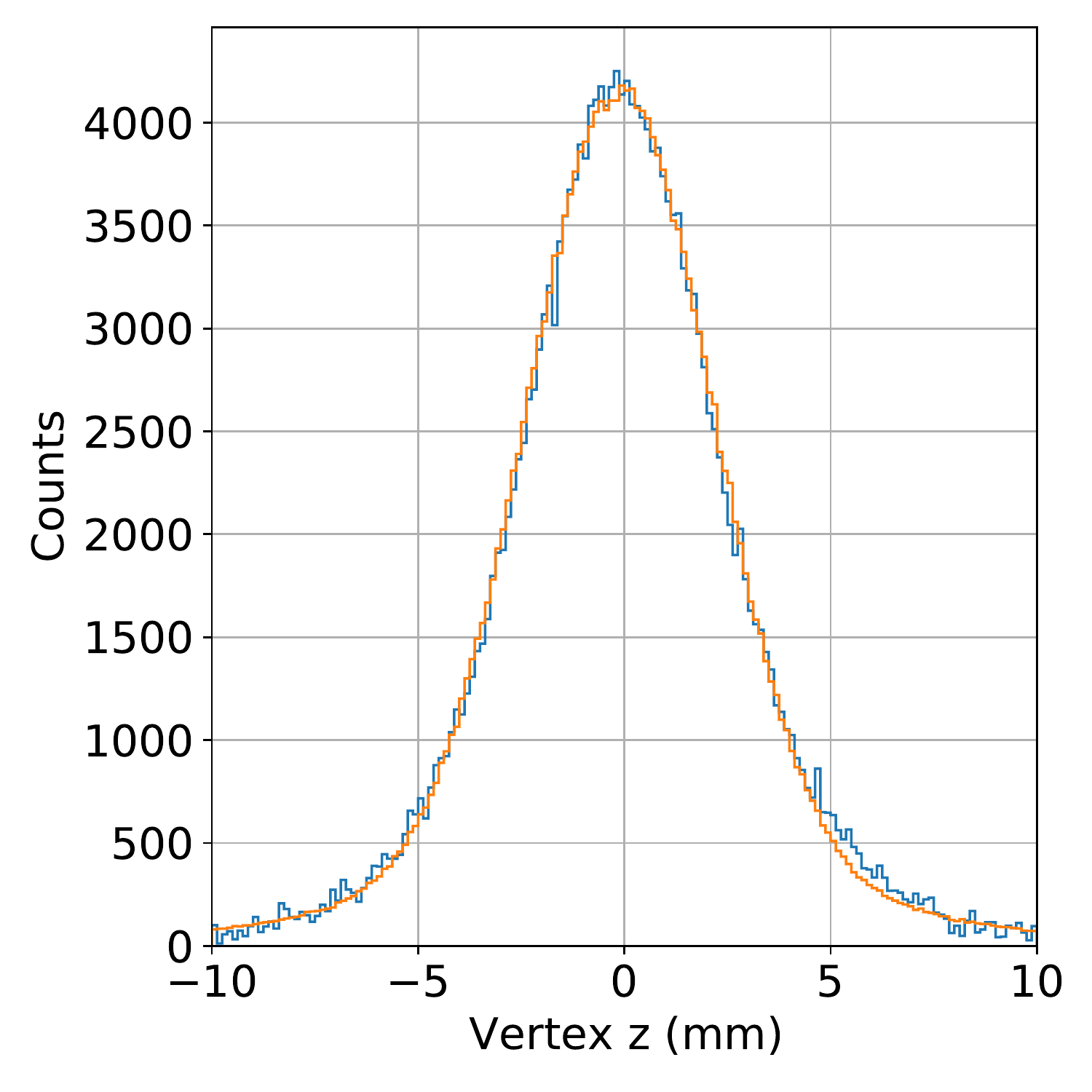}
    \includegraphics[width=0.32\textwidth]{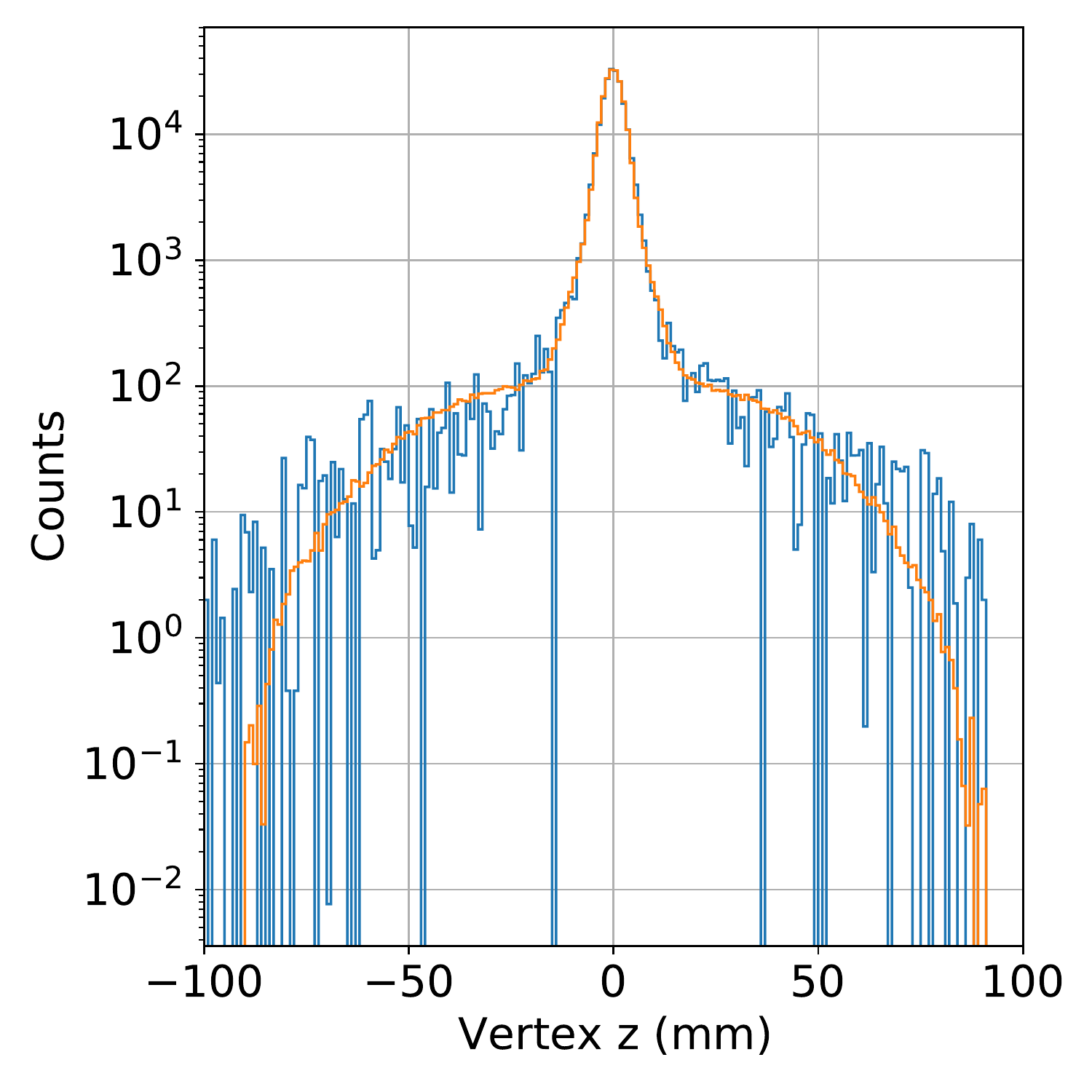}
    \caption{The comparison of the simulation (orange/curve) and extracted signal data (blue/histogram bins) on $\Delta E$, the target coordinates including the in-plane angle $\Phi_0$, out-of-plane angle $\Theta_0$, difference from reference momentum $\Delta P$ and the reconstructed vertex $z$ for the \ang{25} measurement.}
    \label{fig:data_vs_simul}
\end{figure*}

Figure~\ref{fig:data_vs_simul} shows the good agreement between our simulation and data for all the target coordinates. This shows that our correct and appropriate modeling for the radiative process, the target profile, and the resolution of the spectrometer. 

\section{Data analysis}
The data were collected for a total of five weeks over two separate beam times. We measured the elastic electron-proton scattering with $E_\text{beam} = $ \SI{315}{MeV} and scattering angle \ang{15}, \ang{20}, \ang{25}, \ang{30}, and \ang{35} during the first beamtime, \ang{20} and \ang{40} during the second beamtime, covering the four-momentum transfer range $0.01\le Q^2 \le $ \SI{0.065}{(GeV/\clight)^2}.

\subsection{Beamtime summary}

This experiment was the first use of this gas jet target, partially serving as the commissioning of the whole target system. We were able to achieve a maximum flow of \SI{2200}{l_n/h} during the run but we opted for a more conservative flow of \SI{1200}{l_n/h} for most of the beam time to limit the residual pressure inside the scattering chamber, as this helped to limit the backflowing of the residual gas into the beamline towards the accelerator and not to overwhelm the attached turbo-molecular pump. 

The beam halo collimators reduced the background significantly but not completely due to the imperfection of the collimator surface alignment with the beam.
The veto system detected bursts in beam halo with a \SI{50}{Hz} frequency which was previously undetectable by existing instruments. This helped MAMI in identifying a faulty power supply for a correction coil in the accelerator. However, the veto system did not survive the level of vacuum and radiation in the target chamber, so not all background was eliminated.

For a background study, a zero-flow setting should be used. However, running the target with no flow can cause frosting which can block the nozzle, and turning off the cooling can shift the nozzle position due to thermal expansion. Therefore, we collected data at a target flow rate of \SI{50}{l_n/h} for the study of background. 

Further optimization of the pumping power, the nozzle and catcher geometries, and relative positions will help us keep the target running close to the maximum designed flow for prolonged periods in the future, and will also allow true zero-flow background measurements. A new veto detector design and construction to best fit the working condition inside the target chambers are also ongoing. 

\subsection{Data pre-processing}
It takes two steps to translate the raw data we measured, the timing within the VDC, to physics quantities we are interested in, like the electron momentum and scattering angle. These two steps are carefully examined and optimized to ensure the quality of the reconstructed data. 

The first is to reconstruct the particle trajectories at the VDC using the timing information of the wires in the VDC. We noticed false triggering on the scintillator afterpulse and identified these tracks using the maximum VDC drift time measured for each event. We optimized the timing offset and drift velocity by minimizing the track reconstruction uncertainties to compensate for the slight change in gas composition when we replaced the isobutane bottles of the VDCs during the runs. 

The second is to back-propagate the track at the focal plane back to the target coordinates. This step uses a transfer matrix determined by the spectrometer's magnetic field. We performed routine calibration of the parameters in this transfer matrix by measuring electron tracks from a fixed set of known target coordinates limited by the hole on the sieve slit collimator, as shown in Figure~\ref{fig:spec_window}. These calibrated parameters are optimized perturbatively using fits between the simulation and the data to compensate for the possible but tiny shift in the magnet field of the spectrometers between two consecutive calibrations. 

The details of these processes can be found in reference~\cite{wang2021}.

\subsection{Signal extraction}

\begin{figure}
    \centering
    \includegraphics[width=0.45\textwidth]{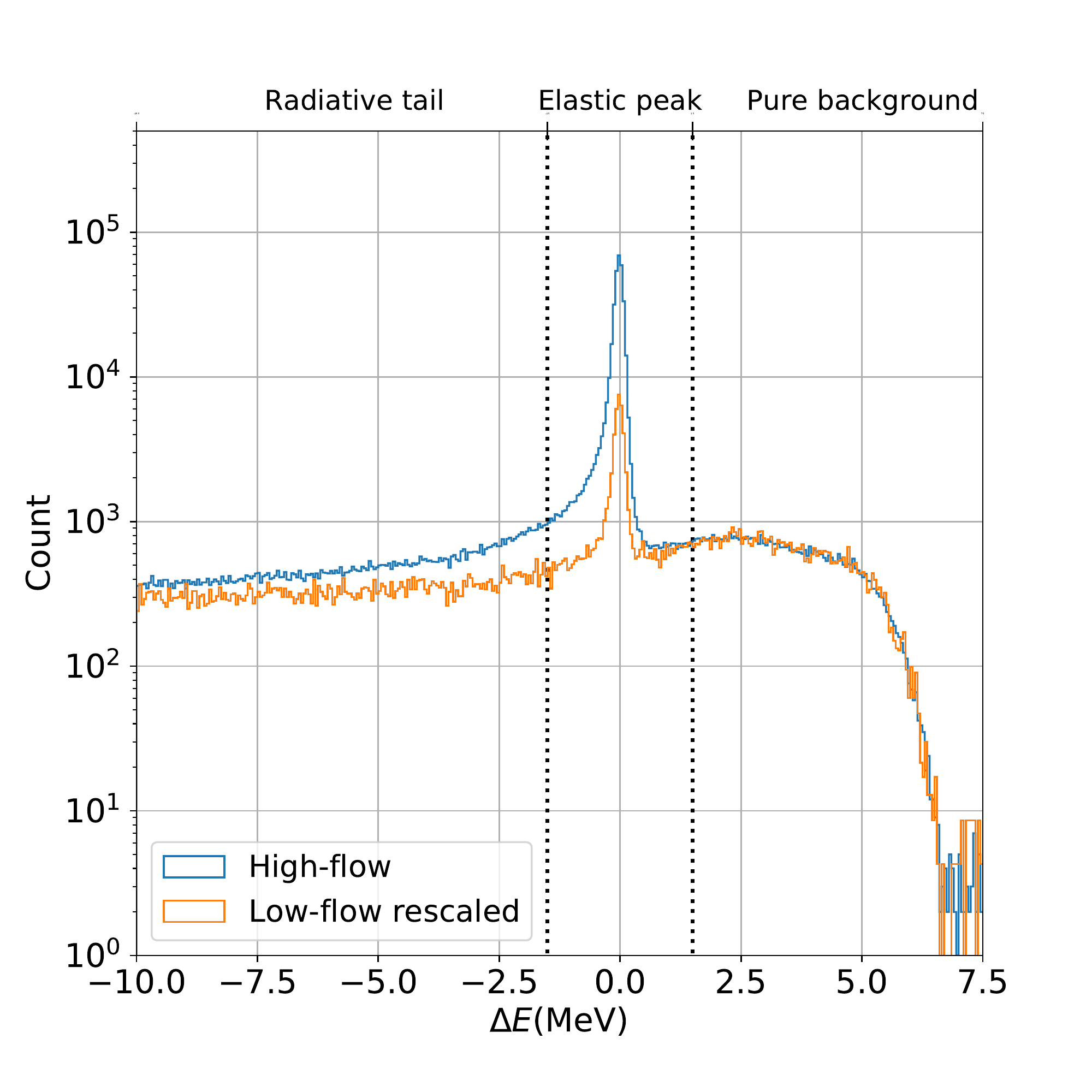}
    \caption{Histogram of $\Delta E$ for high-flow and low-flow settings for the \ang{20} measurement.}
    \label{fig:deltaE_align}
\end{figure}

Figure~\ref{fig:deltaE_align} shows the histogram of the energy difference $\Delta E = E_\text{measured} - E_\text{elastic}(\theta_\text{measured})$ for two different target flows. The elastic peak is well-centered at zero, with the radiation tail extending to negative values. The events with $\Delta E>0$ are purely background events from electrons scattering off heavy nuclei in the target nozzle and catcher. The data from the low-flow setting of \SI{50}{l/h} is re-scaled so that the shapes in the pure background region at positive $\Delta E $ values are aligned. 
We can see from the histogram that the background shape is target-flow independent. 

To determine the total number of elastic events is slightly more complicated since both the high-flow and low-flow data contain elastic electron-proton scattering events, the direct difference from Figure~\ref{fig:deltaE_align} under-counts the total number of signal events. To correctly extract the signal count, we first align the pure background region of the high-flow and low-flow setting data and use the difference to obtain the signal spectra. 
We determine the background spectra by subtracting the re-scaled signal spectra from the high-flow setting data with the scale factor fitted to minimize the variance within the elastic peak window for the background spectra. This process gives us both the background spectra and signal counts in the high-flow data simultaneously. As we mentioned before, Figure~\ref{fig:data_vs_simul} demonstrates the good agreement between our simulation and extracted signal data.

\subsection{Luminosity calculation}
Although we measured the flow rate of the gas jet target in real-time, the exact density distribution that overlaps with the beam profile is not directly measurable. Such fluctuations are beyond the precision we want to achieve. So instead of measuring the total integrated luminosity, we used spectrometer A to measure the electron-proton scattering at \ang{30} during the whole beamtime, as a luminosity monitor. This gives us the relative luminosity for each scattering angle, as measured by spectrometer B, and we can fit the global normalization of the luminosity in the data analysis.

\begin{figure}
    \centering
    \includegraphics[width=0.49\textwidth]{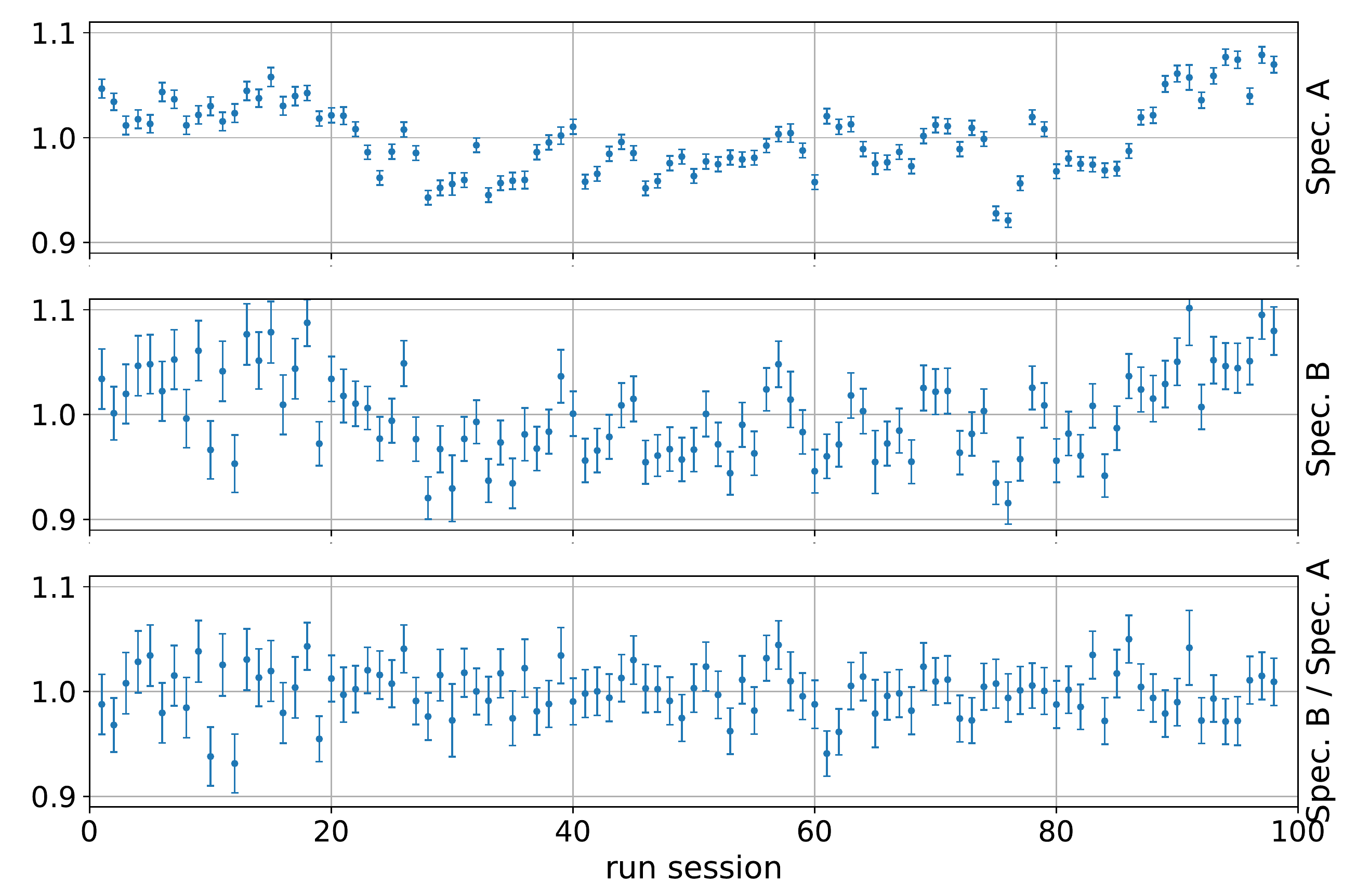}
    \caption{Normalized event rates for spectrometer A, B and their ratio during the \ang{40} measurement. The ratio has $\chi^2=92.0$ for total of 98 sessions. The signal counting rate is about \SI{10}{Hz} in spectrometer A at \ang{30} and \SI{1.1}{Hz} in spectrometer B at \ang{40}.}
    \label{fig:luminosity_monitor}
\end{figure}

Figure~\ref{fig:luminosity_monitor} shows an example of the relative event rate in spectrometers A and B and the ratio of the two for each run session when spectrometer B measures the scattering at \ang{40}. We can see the event rate in each spectrometer has significant fluctuations beyond statistical sources due to the nature of the unstable overlap between the beam and target distribution. As expected, the ratio between two spectrometer events rates is very stable and has only the fluctuation due to statistics with $\chi^2_\text{reduced}$ close to unity for all the other angles as well.

\subsection{Results}
With measurements only at one beam energy, we cannot separate $G_E$ and $G_M$. Instead, we use Kelly's parametrization~\cite{Kelly2004} for $G_M$ and fit the two normalization factors for our measurements, one for each beamtime, based on the PRad's and Mainz's models for $G_E$ measurements~\cite{PRad2019}\cite{Bernauer2014}. The main systematic uncertainties are summarized in Table~\ref{tab:systematic_error}.

\begin{table}[]
\begin{tabular}{ll}
\hline
\hline
Source                             & Uncertainty       \\ \hline
Signal extraction window size      & \textless{}0.17\%  \\
Energy cut in the radiative tail   & \textless{}0.15\% \\
Radiative generator                & \textless{}0.09\%  \\ 
Detector efficiency                & \textless{}0.05\% \\
Magnetic form factor model         & \textless{}0.05\% \\
Cut on the primary vertex location & \textless{}0.03\% \\ \hline
Systematic total                   & \textless{}0.25\% \\ \hline
Statistical (worst at \ang{35})    & \textless{}0.19\% \\
\hline \hline
\end{tabular}
\caption{Summary of the uncertainties.}
\label{tab:systematic_error}
\end{table}

\begin{figure}
    \centering
    \includegraphics[width=0.48\textwidth]{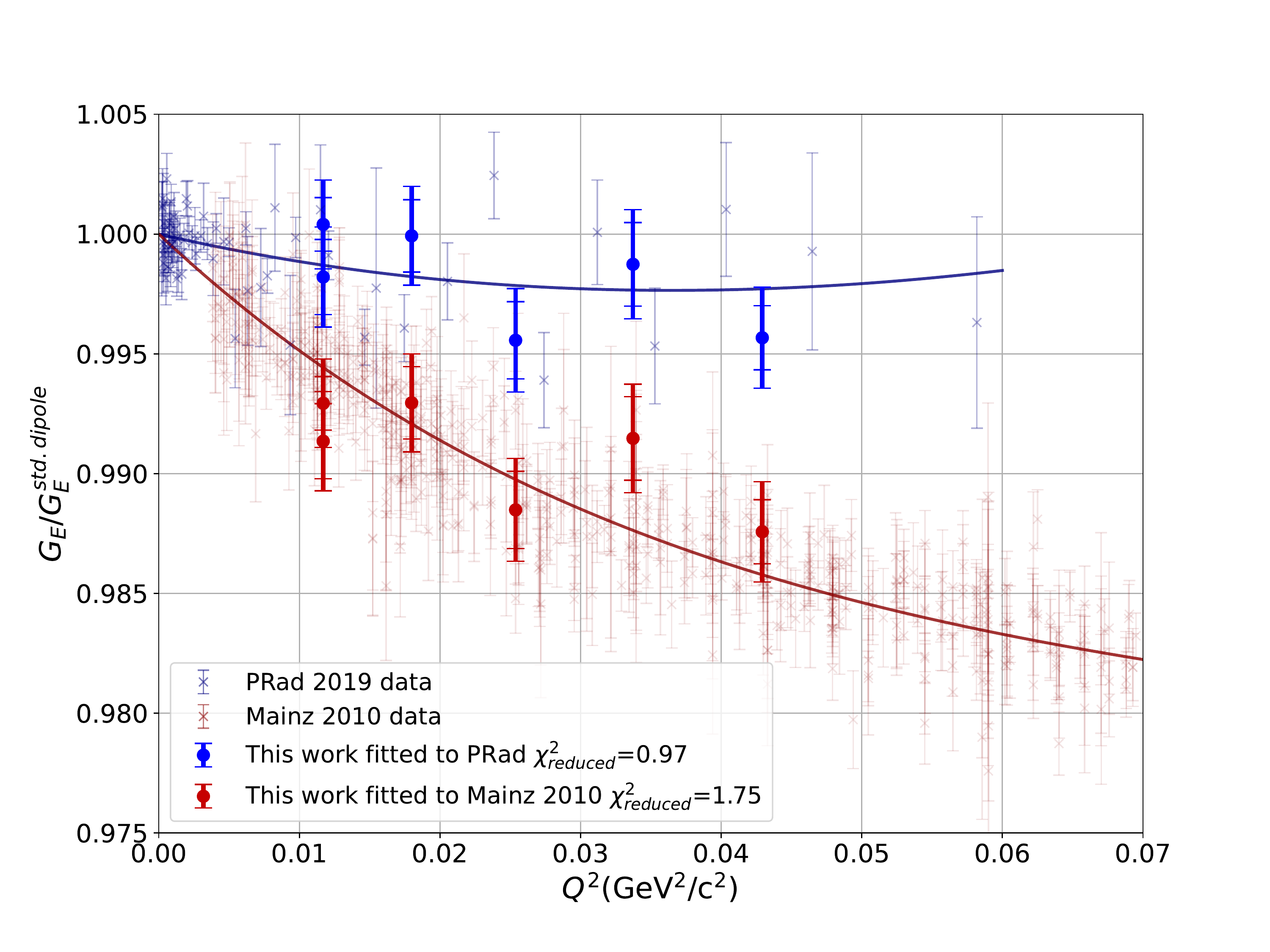}
    \caption{Our measured proton electric form factor $G_E(Q^2)$ with the luminosity normalization factors optimized for the PRad's Rational (1,1) fit ($\chi^2_\text{reduced}$ = 0.97) and Mainz's polynomial fit ($\chi^2_\text{reduced}$ = 1.75).}
    \label{fig:form_factor}
\end{figure}

Figure~\ref{fig:form_factor} shows our result when we fit the global luminosity normalization factors to PRad's Rational (1,1) parametrization~\cite{PRad2019} and Mainz's polynomial parametrization~\cite{Bernauer2014} of the proton electric form factors. With the normalization freedom of the two groups and because of the limited statistics, we find our data consistent with both models and cannot exclude any of them. The data would weakly prefer a slope roughly between the ones indicated by the PRad and Mainz data.

\section{Conclusions}

Our experiment successfully measured the elastic electron-proton scattering within the four-momentum transfer range of $0.01\le Q^2 \le $ \SI{0.045}{(GeV/\clight)^2} using the gas jet target. Our results are consistent with the two recent measurements of the proton electric form factors. However, we cannot discriminate between the two previous measurements due to our limited statistical uncertainty. There is a clear path to improve the precision by optimizing both the jet target itself (subsequent beam times showed more stable operation) and optimization of the collimator-veto system. Our results prove the feasibility of the experiment design using high-resolution spectrometers and the gas jet target for future scattering experiments to resolve the discrepancy in form factor measurements, for example, the MAGIX experiment at MESA\cite{DENIG2016}. 

\section*{Acknowledgments}
We gratefully acknowledge the support from the accelerator group and the technical staff at the
Mainz Microtron. We acknowledge generous support from the U.S. Department of Energy Office of Nuclear Physics under grant DE-FG02-94ER4081 and the U.S. National Science Foundation (NSF) under grant PHY-2012114. 
This work was also supported in part by 
the PRISMA$^+$ (Precision Physics, Fundamental Interactions and
Structure of Matter) Cluster of Excellence,
the Deutsche Forschungsgemeinschaft (DFG, German Research Foundation)
through the Collaborative Research Center 1044 and the Research
Training Group GRK 2128 AccelencE (Accelerator Science and Technology
for Energy-Recovery Linacs),
the Federal State of Rhineland-Palatinate,
the Croatian Science Foundation under Project No. 8570,
and the European Union's Horizon 2020 research and innovation
programme, project STRONG2020, under grant agreement No. 824093.

\end{document}